\documentclass[conference,9pt]{IEEEtran}

\usepackage[english]{babel}								% English language/hyphenation
\makeatletter
\adddialect\l@ENGLISH\l@english
\makeatother

\usepackage[T1]{fontenc}

\usepackage{mathtools}
\usepackage{amssymb, amsthm, amscd, MnSymbol}
\usepackage[inline]{enumitem}

\usepackage{upgreek}
\usepackage{bm}
\usepackage{nicefrac}

\usepackage{float}
\usepackage{stfloats}
\usepackage[section]{placeins}
\usepackage{balance}
\usepackage{cite}
\usepackage{xcolor}

\usepackage[pdftex]{graphicx}	
\usepackage[none]{hyphenat}
\usepackage{widetext}

\usepackage{hyperref} 

\newtheorem{theorem}{Theorem}

\newtheorem{lemma}[theorem]{Lemma}

\newtheorem{corollary}[theorem]{Corollary}

\theoremstyle{remark}
\newtheorem{remark}{Remark}

\newcommand{\X}{\mathbb{X}}

\begin{document}
%\title{T-IT_infGeomBounds}
\title{Generalized Bayesian Cram\'{e}r-Rao Inequality via Information Geometry of Relative $\alpha$-Entropy}

\author{\IEEEauthorblockN{ 
		Kumar Vijay Mishra$^{\dag}$ 
		and M. Ashok Kumar$^{\ddag}$ \\
		{$^{\dag}$}United States CCDC Army Research Laboratory, Adelphi, MD 20783 USA\\
		$^{\ddag}$Department of Mathematics, Indian Institute of Technology Palakkad, 678557 India\\
	Email: kumarvijay-mishra@uiowa.edu, ashokm@iitpkd.ac.in
	}
	}
	
% make the title area
\maketitle

\begin{abstract}
The relative $\alpha$-entropy is the R\'enyi analog of relative entropy and arises prominently in information-theoretic problems. Recent information geometric investigations on this quantity have enabled the generalization of the Cram\'{e}r-Rao inequality, which provides a lower bound for the variance of an estimator of an escort of the underlying parametric probability distribution. However, this framework remains unexamined in the Bayesian framework. In this paper, we propose a general Riemannian metric based on relative $\alpha$-entropy to obtain a generalized Bayesian Cram\'{e}r-Rao inequality. This establishes a lower bound for the variance of an unbiased estimator for the $\alpha$-escort distribution starting from an unbiased estimator for the underlying distribution. We show that in the limiting case when the entropy order approaches unity, this framework reduces to the conventional Bayesian Cram\'{e}r-Rao inequality. Further, in the absence of priors, the same framework yields the deterministic Cram\'{e}r-Rao inequality.
\end{abstract}

\begin{IEEEkeywords}
Bayesian bounds, cross-entropy, R\'enyi entropy, Riemannian metric, Sundaresan divergence.
\end{IEEEkeywords}

\IEEEpeerreviewmaketitle

\section{Introduction}
\label{sec:intro}
In information geometry, a parameterized family of probability distributions is expressed as a manifold in the Riemannian space \cite{spivak2005comprehensive}, in which the parameters form the coordinate system on manifold and the distance measure is given by the Fisher information matrix (FIM) \cite{gallot2004riemannian}. This framework reduces certain important information-theoretic problems to investigations of different Riemannian manifolds \cite{ay2017information}. This perspective is helpful in analyzing many problems in engineering and sciences where probability distributions are used, including optimization \cite{amari2013minkovskian}, signal processing \cite{amari2016information}, machine learning \cite{amari1998natural}, optimal transport \cite{gangbo1996geometry}, and quantum information \cite{grasselli2001uniqueness}. 

In particular, when the separation between the two points on the manifold is defined by \textit{Kullback-Leibler divergence} (KLD) or \textit{relative entropy}
between two probability distributions $p$ and $q$ on a finite state space $\X = \{0,1,2,\dots,M\}$, i.e.,
	\begin{align}
	    \mathcal{I}(p,q) := \sum_{x\in\mathbb{X}} p(x)\log\frac{p(x)}{q(x)},
	\end{align}
then the resulting \textit{Riemmanian metric} is defined by FIM \cite{amari2000methods}. This method of defining a Riemannian metric on statistical manifolds from a general divergence function is due to Eguchi \cite{1992xxHMJ_Egu}. Since FIM is the inverse of the well-known \textit{deterministic} Cram\'{e}r-Rao lower bound (CRLB), the information-geometric results are directly connected with those of estimation theory. Further, the relative entropy is related to the Shannon entropy $H(p) := -\sum_{x\in\mathbb{X}} p(x)\log p(x)$ by $\mathcal{I}(p,u) = \log |\mathbb{X}| - H(p)$, where $u$ is the uniform distribution on $\mathbb{X}$. 

It is, therefore, instructive to explore information-geometric frameworks for key estimation-theoretic results. For example, the Bayesian CRLB \cite{van2013detection,gill1995applications} is the analogous lower bound to CRLB for random variables. It assumes the parameters to be random with an \textit{a priori} probability density function.
In \cite{kumar2018information}, we derived Bayesian CRLB using a general definition of KLD when the probability densities are not normalized. 

Recently, \cite{kumar2020cram} studies information geometry of \textit{R\'{e}nyi entropy} \cite{renyi1961measures}, which is a generalization of Shannon entropy. In source coding problem where normalized cumulants of compressed lengths are considered instead of expected compressed lengths, R\'enyi entropy is used as a measure of uncertainty \cite{1965xxIC_Cam}. The R\'{e}nyi entropy of $p$ of order $\alpha$, $\alpha \ge 0$, $\alpha\neq 1$, is defined to be $H_{\alpha}(p) := \frac{1}{1-\alpha}\log\sum_xp(x)^{\alpha}$. In the context of source distribution version of this problem, the R\'{e}nyi analog of relative entropy is \textit{relative $\alpha$-entropy} \cite{198809TIT_BluMcE,200701TIT_Sun}. The relative $\alpha$-entropy of $p$ with respect to $q$ (or \textit{Sundaresan's divergence} between $p$ and $q$) is defined as
	\begin{align}
	\label{eqn:alphadiv}
      \mathcal{I}_{\alpha}(p,q)	:=& \frac{\alpha}{1-\alpha} \log \sum_x p(x) q(x)^{\alpha-1} \nonumber\\
      &- \frac{1}{1-\alpha}\log \sum_x p(x)^{\alpha} + \log \sum_x q(x)^{\alpha}.
	\end{align}
	 It follows that, as $\alpha \rightarrow 1$, we have $\mathcal{I}_{\alpha}(p,q) \rightarrow \mathcal{I}(p\|q)$ %, the relative entropy of $p$ with respect to $q$, 
	 and $H_{\alpha}(p) \rightarrow H(p)$ \cite{kumar2015minimization-1}. R\'enyi entropy and relative $\alpha$-entropy are related by the equation $\mathcal{I}_\alpha(p,u) = \log |\mathbb{X}| - H_\alpha(p)$. 
	Relative $\alpha$-entropy is closely related to the \textit{Csisz\'ar $f$-divergence} $D_{f}$ as
	\begin{eqnarray}
    \label{eqn:alphadiv2}
    \mathcal{I}_{\alpha}(p,q) = \frac{\alpha}{1-\alpha} \log\left[ \text{sgn}(1-\alpha) \cdot D_{f}(p^{(\alpha)},q^{(\alpha)}) + 1\right],
    \end{eqnarray}
    %\vspace*{-0.4cm}
where $p^{(\alpha)}(x) := \frac{p(x)^{\alpha}}{\sum_y {p(y)}^{\alpha}}$, $q^{(\alpha)}(x) := \frac{q(x)^{\alpha}}{\sum_y {q(y)}^{\alpha}}$, and $f(u) = \textrm{sgn}(1-\alpha) \cdot (u^{{1}/{\alpha}} - 1), u \geq 0$ \cite[Sec.~II]{kumar2015minimization-1}. The measures $p^{(\alpha)}$ and $q^{(\alpha)}$ are called {\em $\alpha$-escort} or {\em $\alpha$-scaled} measures \cite{1998xxPhyA_Tsa,201802NCC_KarSun}. It is easy to show that, indeed, the right side of (\ref{eqn:alphadiv2}) is the R\'enyi divergence between $p^{(\alpha)}$ and $q^{(\alpha)}$ of order ${1}/{\alpha}$. 
	 
The R\'enyi entropy and relative $\alpha$-entropy arise in several important information-theoretic problems such as guessing \cite{199601TIT_Ari,200701TIT_Sun,Huleihel} and task encoding  \cite{2014xxarx_BunLap}. Relative $\alpha$-entropy arises in statistics as a generalized likelihood function robust to outliers \cite{2001xxBio_Jon_etal}, \cite{kumar2015minimization-2}. It also shares many interesting properties with relative entropy; see, e.g. \cite[Sec.~II]{kumar2015minimization-1} for a summary. For example, relative $\alpha$-entropy behaves like squared Euclidean distance and satisfies a Pythagorean property in a similar way relative entropy does \cite{kumar2015minimization-1,kumar2018information}. This property helps in establishing a computation method \cite{kumar2015minimization-2} for a robust estimation procedure \cite{2008xxJma_FujEgu}. 

Motivated by such analogous relationships, our previous works \cite{kumar2020cram} investigated the relative $\alpha$-entropy from a differential geometric perspective. In particular, we applied Eguchi's method with relative entropy as the divergence function to obtain the resulting statistical manifold with a general Riemannian metric. This metric is specified by the Fisher information matrix that is the inverse of the so called \textit{deterministic $\alpha$-CRLB} \cite{kumar2015minimization-1}. In this paper, we study the structure of statistical manifolds with respect to a relative $\alpha$-entropy in a Bayesian setting. This is a non-trivial extension of our work in \cite{kumar2018information}, where we proposed Riemmanian metric arising from the relative entropy for the Bayesian case. In the process, we derive a \textit{general Bayesian Cram\'{e}r-Rao inequality} and the resulting \textit{Bayesian $\alpha$-CRLB} which embed the compounded effects of both R\'{e}nyi order $\alpha$ and Bayesian prior distribution. We show that, in limiting cases, the bound reduces to deterministic $\alpha$-CRLB (in the absence of prior), Bayesian CRLB (when $\alpha \rightarrow 1$) or CRLB (no priors and $\alpha \rightarrow 1$).

The rest of the paper is organized as follows. In the next section, we provide the essential background to information geometry. We then introduce the definition of Bayesian relative $\alpha$-entropy in Section~\ref{sec:relalphaBayes} and show that it is a valid divergence function. In Section \ref{sec:fim}, we establish the connection between this divergence and the Riemannian metric and then derive the Bayesian $\alpha$-version of Cram\'{e}r-Rao inequality in Section~\ref{sec:analogous_cr_inequality}. Finally, we state our main result for the Bayesian $\alpha$-CRLB in Section \ref{sec:error_bounds} and conclude in Section \ref{sec:conc}.

\section{Desiderata for Information Geometry}
\label{sec:inf}
A $n$-dimensional manifold is a Hausdorff and second countable topological space which is locally homeomorphic to Euclidean space of dimension $n$ \cite{gallot2004riemannian}. A Riemannian manifold is a real differentiable manifold in which the tangent space at each point is a finite dimensional Hilbert space and, therefore, equipped with an inner product. The collection of all these inner products is Riemannian metric. In information geometry, the statistical models play the role of a manifold and the Fisher information matrix and its various generalizations play the role of a Riemannian metric. The statistical manifold here means a parametric family of probability distributions $S = \{p_\theta: \theta\in\Theta\}$ with a continuously varying parameter space $\Theta$ (statistical model). The dimension of a statistical manifold is the dimension of the parameter space. For example, $S = \{N(\mu,\sigma^2) : \mu\in \mathbb{R}, \sigma^2 > 0\}$ is a two dimensional statistical manifold. The tangent space at a point of $S$ is a linear space that corresponds to a ``local linearization'' at that point. The tangent space at a point $p$ of $S$ is denoted by $T_{\tilde{p}}(S)$. The elements of $T_{\tilde{p}}(S)$ are called {\em tangent vectors} of $S$ at $p$. A Riemannian metric at point $p$ of $S$ is an inner product defined for any pair of tangent vectors of $S$ at $p$.

Let us restrict to statistical manifolds defined on a finite set $\mathcal{X} = \{a_1,\dots,a_d\}$.  Let $\mathcal{P} := \mathcal{P}(\mathcal{X})$ denote the space of all probability distributions on $\mathcal{X}$. Let $S\subset\mathcal{P}$ be a sub-manifold. Let $\theta = (\theta_1,\dots,\theta_k)$ be a parameterization of $S$. By a {\em divergence}, we mean a non-negative function $D$ defined on $S\times S$ such that $D(p,q) = 0$ iff $p=q$. Given a divergence function on $S$, Eguchi \cite{eguchi1992geometry} defines a Riemannian metric on $S$ by the matrix
\[
G^{(D)}(\theta) = \left[g_{i,j}^{(D)}(\theta)\right],
\]
where
\begin{eqnarray*}
	g_{i,j}^{(D)}(\theta) := -D[\partial_i,\partial_j] :=  -\frac{\partial}{\partial\theta_j'}\frac{\partial}{\partial\theta_i}D(p_{\theta},p_{\theta'})\bigg|_{\theta=\theta'}
\end{eqnarray*}
where $g_{i,j}$ is the elements in the $i$th row and $j$th column of the matrix $G$, $\theta = (\theta_1,\dots,\theta_n)$, $\theta' = (\theta_1',\dots,\theta_n')$, and dual affine connections $\nabla^{(D)}$ and $\nabla^{(D^*)}$, with connection coefficients described by following Christoffel symbols
\begin{eqnarray*}
	\Gamma_{ij,k}^{(D)}(\theta) := -D[\partial_i\partial_j,\partial_k]
	:= -\frac{\partial}{\partial\theta_i}\frac{\partial}{\partial\theta_j}\frac{\partial}{\partial\theta_k'}D(p_{\theta},p_{\theta'})\bigg|_{\theta=\theta'}
\end{eqnarray*}
and
\begin{eqnarray*}
	\Gamma_{ij,k}^{(D^*)}(\theta) & := & -D[\partial_k,\partial_i\partial_j] :=  -\frac{\partial}{\partial\theta_k}\frac{\partial}{\partial\theta_i'}\frac{\partial}{\partial\theta_j'}D(p_{\theta},p_{\theta'})\bigg|_{\theta=\theta'},
\end{eqnarray*}
such that, $\nabla^{(D)}$ and $\nabla^{(D^*)}$ form a dualistic structure in the sense that
\begin{eqnarray}
	\label{dualistic-structure}
	\partial_k g_{i,j}^{(D)}=\Gamma_{ki,j}^{(D)}+ \Gamma_{kj,i}^{(D^*)},
\end{eqnarray}
where $D^*(p,q) = D(q,p)$.

\section{Relative \texorpdfstring{$\alpha$}{}-entropy in the Bayesian Setting}
\label{sec:relalphaBayes}
We now introduce relative $\alpha$-entropy in the Bayesian case. Define $S = \{p_\theta: \theta = (\theta_1,\dots,\theta_k)\in\Theta\}$ as a $k$-dimensional sub-manifold of $\mathcal{P}$ and
\begin{eqnarray}
\label{eqn:denormalized_manifold}
\tilde{S} := \{\tilde{p}_{\theta}(x) = p_{\theta}(x)\lambda(\theta) : p_{\theta}\in S\},
\end{eqnarray}
where $\lambda$ is a probability distribution on $\Theta$. Then, $\tilde{S}$ is a sub-manifold of $\tilde{\mathcal{P}}$. Let $\tilde{p}_\theta, \tilde{p}_{\theta'}\in \tilde{S}$. The relative entropy of $\tilde{p}_\theta$ with respect to $\tilde{p}_{\theta'}$ is (c.f. \cite[Eq.~ (2.4)]{1991xxTAS_Csi} and \cite{kumar2018information})
\begin{eqnarray}
	\label{eqn:KL-div}
	I(\tilde{p}_\theta\|\tilde{p}_{\theta'}) & = & \sum_x \tilde{p}_{\theta}(x) \log \frac{\tilde{p}_{\theta}(x)}{\tilde{p}_{\theta'}(x)} - \sum_x \tilde{p}_{\theta}(x) + \sum_x \tilde{p}_{\theta'}(x)\nonumber\\
	& = & \sum_x p_{\theta}(x)\lambda(\theta) \log \frac{p_{\theta}(x)\lambda(\theta)}{p_{\theta'}(x)\lambda(\theta')} - \lambda(\theta) + \lambda(\theta').\nonumber
\end{eqnarray}

We define relative $\alpha$-entropy of $\tilde{p}_\theta$ with respect to $\tilde{p}_{\theta'}$ by
\begin{eqnarray}
 \lefteqn{\mathcal{I}_{\alpha}(\tilde{p}_{\theta},\tilde{p}_{\theta'})}\nonumber\\
    & \hspace{-0.65cm} := \frac{\lambda(\theta)}{1-\alpha}\log\sum_x p_\theta(x) (\lambda(\theta')p_{\theta'}(x))^{\alpha-1}+ \lambda(\theta')\nonumber\\
&  \hspace{-0.3cm} - \lambda(\theta)\left[\frac{\log\sum_xp_\theta(x)^\alpha}{\alpha (1-\alpha)} - \{1 + \log \lambda(\theta)\} -\frac{1}{\alpha} \log\sum_xp_{\theta'}(x)^\alpha\right]\nonumber
\end{eqnarray}

We present the following Lemma~\ref{lem:rel} which shows that our definition of Bayesian relative $\alpha$-entropy is not only a valid divergence function but also coincides with the KLD as $\alpha\to 1$.
\begin{lemma}
\label{lem:rel}
\begin{enumerate}
\item[]
\item $\mathcal{I}_{\alpha}(\tilde{p}_{\theta},\tilde{p}_{\theta'})\ge 0$ with equality if and only if $\tilde{p}_{\theta} = \tilde{p}_{\theta'}$
\item $\mathcal{I}_{\alpha}(\tilde{p}_{\theta},\tilde{p}_{\theta'})\to I(\tilde{p}_\theta\|\tilde{p}_{\theta'})$ as $\alpha\to 1$.
\end{enumerate}

\end{lemma}
\begin{IEEEproof}
1) Let $\alpha >1$. Applying Holder's inequality with Holder conjugates $p=\alpha$ and $q={\alpha}/{(\alpha-1)}$, we have
    \[
    \sum_x p_\theta(x) (\lambda(\theta')p_{\theta'}(x))^{\alpha-1} \le \|p_\theta\| \lambda(\theta')^{\alpha-1}\|p_\theta'\|^{\alpha-1},
    \]
    where $\|\cdot\|$ denotes $\alpha$-norm. When $\alpha <1$, the inequality is reversed. Hence
    \begin{flalign*}
        &\lefteqn{\frac{\lambda(\theta)}{1-\alpha}\log\sum_x p_\theta(x) (\lambda(\theta')p_{\theta'}(x))^{s-1}}\nonumber\\
        &\ge \frac{\lambda(\theta)\log\sum_xp_\theta(x)^\alpha}{\alpha(1-\alpha)} - \lambda(\theta)\log \lambda(\theta') - \frac{\lambda(\theta)}{\alpha}\log\sum_xp_{\theta'}(x)^\alpha\nonumber\\
& \ge \frac{\lambda(\theta)\log\sum_xp_\theta(x)^\alpha}{\alpha(1-\alpha)} - \lambda(\theta)\log \lambda(\theta) - \lambda(\theta) + \lambda(\theta')\nonumber\\
& \hfill - \frac{\lambda(\theta)}{\alpha}\log\sum_xp_{\theta'}(x)^\alpha\\
&= \lambda(\theta)\left[\frac{\log\sum_xp_\theta(x)^\alpha}{\alpha(1-\alpha)} - \{1 + \log \lambda(\theta)\} - \log\sum_xp_{\theta'}(x)^\alpha\right] + \lambda(\theta'),\nonumber
\end{flalign*}
where the second inequality follows because, for $x,y\ge 0$,
\begin{align*}
\log\frac{x}{y} = -x\log\frac{y}{x} \ge -x (y/x-1) \ge -y + x,
\end{align*}
and hence $$x\log y \le x\log x - x + y.$$
The conditions of equality follow from the same in Holder's inequality and $\log x\le x-1$.

  2) This follows by applying L'H\^{o}pital rule to the first term of $I_\alpha$:
\begin{flalign}
    &\lim_{\alpha\to 1}\left[\frac{\alpha}{1-\alpha} \lambda(\theta) \log \sum_x p_{\theta}(x) (\lambda(\theta')p_{\theta'}(x))^{\alpha-1}\right]\nonumber\\
    &=\lim_{\alpha\to 1}\left[\frac{1}{\frac{1}{\alpha}-1} \lambda(\theta) \log \sum_x p_{\theta}(x) (\lambda(\theta')p_{\theta'}(x))^{\alpha-1}\right]\nonumber\\
    &=\lambda(\theta)\lim_{\alpha\to 1}\left[\frac{1}{-\frac{1}{\alpha^2}}\frac{\sum_x p_{\theta}(x) (\lambda(\theta')p_{\theta'}(x))^{\alpha-1} \log(\lambda(\theta')p_{\theta'}(x))}{\sum_x p_{\theta}(x) (\lambda(\theta')p_{\theta'}(x))^{\alpha-1}}\right]\nonumber\\
    &= -\sum_x (\lambda(\theta)p_{\theta}(x)) \log(\lambda(\theta')p_{\theta'}(x)),\nonumber
\end{flalign}
and since Renyi entropy coincides with Shannon entropy as $\alpha\to 1$.
\end{IEEEproof}

\section{Fisher Information Matrix for the Bayesian Case}
\label{sec:fim}
The Eguchi's theory we provided in section \ref{sec:inf} can also be extended to the space $\tilde{\mathcal{P}}(\mathcal{X})$ of all positive measures on $\mathcal{X}$, that is, $\tilde{\mathcal{P}} = \{\tilde{p}:\mathcal{X}\to (0,\infty)\}$. 
Following Eguchi \cite{eguchi1992geometry}, we define a Riemannian metric
$[g_{i,j}^{(I_{\alpha})}(\theta)]$ on $\tilde{S}$ by
\begin{widetext}
\begin{flalign}
	&{g_{i,j}^{(I_{\alpha})}(\theta)}\nonumber\\
	&=-\frac{\partial}{\partial\theta_j'}\frac{\partial}{\partial\theta_i}I_{\alpha}(\tilde{p}_{\theta},\tilde{p}_{\theta'})\bigg|_{\theta' = \theta}\nonumber\\
	&=  \frac{1}{\alpha-1}\cdot\partial_j'\partial_i \lambda(\theta)\log \sum_y p_{\theta}(x) ({\lambda(\theta') p_{\theta'}(x)})^{\alpha-1}\bigg|_{\theta' = \theta} - \partial_i\lambda(\theta)\partial_j'\log\sum_x p_{\theta'}(x)^\alpha\bigg|_{\theta' = \theta}\nonumber\\
	\label{eqn:alpha-metric-bayesian}
    & = \frac{1}{\alpha-1}\left\{\lambda(\theta)\sum_x \partial_i p_{\theta}(x)\cdot \partial_j'\left[\frac{({\lambda(\theta') p_{\theta'}(x)})^{\alpha-1}}{\sum_y p_{\theta}(y) ({\lambda(\theta') p_{\theta'}(x)})^{\alpha-1}}\right]_{\theta' = \theta} + \partial_i\lambda(\theta)\cdot \left[\frac{{\sum_x p_\theta(x) \partial_j'\big(\lambda(\theta)p_{\theta'}(x)}\big)^{\alpha-1}}{\sum_x p_{\theta}(x) ({\lambda(\theta') p_{\theta'}(x)})^{\alpha-1}}\right]_{\theta' = \theta}\right\}\nonumber\\
    &\hspace{8cm} - \partial_i\lambda(\theta)\partial_j'\log\sum_xp_{\theta'}(x)^\alpha\bigg|_{\theta' = \theta}
	\\
	& = \lambda(\theta) \left\{ \frac{\sum_x\partial_i p_{\theta}(x){(\lambda(\theta)p_{\theta}(x)})^{\alpha-2}\partial_j (\lambda(\theta)p_{\theta}(x))}{\sum_x p_{\theta}(x)(\lambda(\theta)p_{\theta}(x))^{\alpha-1}} - \frac{\sum_x (\partial_i p_{\theta}(x)){p_{\theta}(x)}^{\alpha-1}}{\sum_x p_{\theta}(x)^{\alpha}}
	\cdot \frac{\sum_x p_\theta(x) (\lambda(\theta)p_{\theta'}(x))^{\alpha-2}\partial_j(\lambda(\theta)p_{\theta}(x))}{\sum_x p_{\theta}(x) ({\lambda(\theta) p_{\theta}(x)})^{\alpha-1}} \right.\nonumber\\
	&\;\;\;\;\;\;\;\;\;\;\;\;\;\;\;\;\;\;\left.+ \partial_i\log\lambda(\theta)\cdot \left[\frac{{\sum_x p_\theta(x) \partial_j'\big(\lambda(\theta)p_{\theta'}(x)}\big)^{\alpha-1}}{\sum_x p_{\theta}(x) ({\lambda(\theta') p_{\theta'}(x)})^{\alpha-1}}\right]_{\theta' = \theta}\right\} - \partial_i\lambda(\theta)E_{\theta^{(\alpha)}}[\partial_j \log p_{\theta}(X)]
	\nonumber\\
	& = \lambda(\theta) \left\{E_{\theta^{(\alpha)}}[\partial_i \log p_{\theta}(X)\partial_j \log p_{\theta}(X)] + \partial_j\log\lambda(\theta)E_{\theta^{(\alpha)}}[\partial_i \log p_{\theta}(X)] - E_{\theta^{(\alpha)}}[\partial_i \log p_{\theta}(X)]\left[E_{\theta^{(\alpha)}}[\partial_j \log p_{\theta}(X)]+ \partial_j\log\lambda(\theta)\right]\right.\nonumber\\
	&\left.\;\;\;\;\;\;\;\;\;\;\;\;\;\;\;\;\;\;\;\;\;\;\;\;\;\;\;\;\;\;\;\;\;\;\;\;\;\;\;\;\;\;\;\;\;\;\;\;\;\;\;\;\;\;\;\;\;\;\;\;\;\;\;\;\;\;\;\;\;\;\;\;  + \partial_i\log\lambda(\theta)\cdot \left[E_{\theta^{(\alpha)}}[\partial_j \log p_{\theta}(X)] +  {\partial_j\log\lambda(\theta)}\right]\right\} - \partial_i\lambda(\theta)E_{\theta^{(\alpha)}}[\partial_j \log p_{\theta}(X)]\nonumber\\
	& = \lambda(\theta) \left[\text{Cov}_{\theta^{(\alpha)}}[\partial_i \log p_{\theta}(X), \partial_j \log p_{\theta}(X)] + \partial_i\log\lambda(\theta)\cdot \{E_{\theta^{(\alpha)}}[\partial_j \log p_{\theta}(X)] + \partial_j\log\lambda(\theta)\}\right] - \partial_i\lambda(\theta)E_{\theta^{(\alpha)}}[\partial_j \log p_{\theta}(X)]\nonumber\\
	& = \lambda(\theta) \left\{\text{Cov}_{\theta^{(\alpha)}}[\partial_i \log p_{\theta}(X), \partial_j \log p_{\theta}(X)] + \partial_i\log\lambda(\theta)\partial_j\log\lambda(\theta)\right\}\nonumber\\
	& = \lambda(\theta)[g_{i,j}^{(\alpha)}(\theta) + J_{i,j}^{\lambda}(\theta)],
\end{flalign}
\end{widetext}
where
\begin{equation}
\label{eqn:Fisher_metric}
g_{i,j}^{(\alpha)}(\theta) := \text{Cov}_{\theta^{(\alpha)}}[\partial_i \log p_{\theta}(X), \partial_j \log p_{\theta}(X)],
\end{equation}
and
\begin{equation}
\label{eqn:J_matrix}
J_{i,j}^{\lambda}(\theta) := \partial_i (\log \lambda(\theta))\cdot\partial_j (\log \lambda(\theta)).
\end{equation} 
Let $G^{(\alpha)}(\theta) := [g^{(\alpha)}_{i,j}(\theta)]$, $J^{\lambda}(\theta) := [J_{i,j}^{\lambda}(\theta)]$ and 
$G_\alpha^{\lambda}(\theta) := G^{(\alpha)}(\theta) + J^{\lambda}(\theta)$.
Notice that, when $\alpha = 1$, $G_\alpha^{\lambda}$ becomes $G^{(I)}$, the usual Fisher information matrix in the Bayesian case [c.f. \cite{kumar2018information}].

\section{An \texorpdfstring{$\alpha$}{}-Version of Cram\'{e}r-Rao Inequality in the Bayesian Setting}
	\label{sec:analogous_cr_inequality}
	We now investigate the geometry of $\tilde{\mathcal{P}}$ with respect to the metric $G_\alpha^{\lambda}$. Later, we formulate %come up with 
	an $\alpha$-equivalent version of the Cram\'{e}r-Rao inequality associated with a submanifold $\tilde{S}$.
	Observe that $\tilde{\mathcal{P}}$ is a subset of $\mathbb{R}^{\tilde{\mathcal{X}}}$, where $\tilde{\mathcal{X}} := \mathcal{X}\cup\{a_{d+1}\}$. The tangent space at every point of $\tilde{\mathcal{P}}$ is $\mathcal{A}_0 := \{A\in\mathbb{R}^{\tilde{\mathcal{X}}} : \sum_{x\in\tilde{\mathcal{X}}}A(x) = 0\}$. That is, $T_{\tilde{p}}(\tilde{\mathcal{P}}) = \mathcal{A}_0$. We denote a tangent vector (that is, elements of $\mathcal{A}_0$) by $X^{(m)}$. The manifold $\tilde{\mathcal{P}}$ can be recognized by its homeomorphic image $\{\log \tilde{p}: \tilde{p}\in\tilde{\mathcal{P}}\}$ under the mapping $\tilde{p}\mapsto \log \tilde{p}$. Under this mapping the tangent vector $X\in T_{\tilde{p}}(\tilde{\mathcal{P}})$ can be represented $X^{(e)}$ which is defined by $X^{(e)}(x) = X^{(m)}(x)/\tilde{p}(x)$ and we define
\begin{equation}
\label{exponential_tangent_space}
	T_{\tilde{p}}^{(e)}(\tilde{\mathcal{P}}) = \{X^{(e)} : X\in T_{\tilde{p}}(\tilde{\mathcal{P}})\} = \{A\in\mathbb{R}^{\tilde{\mathcal{X}}} : \mathbb{E}_{\tilde{p}}[A] = 0\}.
\end{equation}

Motivated by the expression for the Riemannian metric in (\ref{eqn:alpha-metric-bayesian}), define
	\begin{flalign}
	\label{eqn:alpha_representation}
	&\lefteqn{\partial_i^{(\alpha)}(p_\theta (x))}\nonumber\\
	& := \frac{1}{\alpha-1}\partial_i'\left(\frac{{p_{\theta'}(x)}^{\alpha-1}}{\sum_y p_{\theta}(y)\, {p_{\theta'}(y)}^{\alpha-1}}\right)\bigg |_{\theta'=\theta}\nonumber\\
	&  = \frac{1}{\alpha-1}\partial_i'\left(\frac{{p_{\theta'}(x)}^{\alpha-1}}{\sum_y p_{\theta}(y) {p_{\theta'}(y)}^{\alpha-1}}\right)\bigg |_{\theta'=\theta}\nonumber\\
	& = \left[\frac{{p_{\theta}(x)}^{\alpha-2}~\partial_i p_{\theta}(x)}{\sum_y {p_{\theta}(y)}^{\alpha}} - \frac{{p_{\theta}(x)}^{\alpha-1}~\sum_y {p_{\theta}(y)}^{\alpha-1}\partial_i p_{\theta}(y)}{(\sum_y {p_{\theta}(y)}^{\alpha})^2}\right]\nonumber\\
	& = \left[\frac{{p_{\theta}(x)}^{(\alpha)}}{p_{\theta}(x)}\partial_i(\log p_{\theta}(x)) - \frac{{p_{\theta}(x)}^{(\alpha)}}{p_{\theta}(x)}E_{\theta^{(\alpha)}}[\partial_i(\log p_{\theta}(X))]\right].\nonumber\\
	\end{flalign}
	We shall call the above an \emph{$\alpha$-representation of $\partial_i$ at $p_\theta$}. With this notation, the $G^{(\alpha)}$ is given by
	\begin{equation*}
	g_{i,j}^{(\alpha)}(\theta) = \sum_x \partial_i p_{\theta}(x) \cdot \partial_i^{(\alpha)}(p_\theta(x)).
	\end{equation*}
	It should be noted that $E_{\theta}[\partial_i^{(\alpha)}(p_\theta(X))] = 0$. This follows since %is obvious once we check that
	\[
	\partial_i^{(\alpha)} (p_{\theta}) = \frac{p_{\theta}^{(\alpha)}}{p_{\theta}} \partial_i \log p_{\theta}^{(\alpha)}.
	\]
	When $\alpha =1$, the right hand side of (\ref{eqn:alpha_representation}) reduces to $\partial_i(\log p_{\theta})$.
	
	Motivated by (\ref{eqn:alpha_representation}), the \emph{$\alpha$-representation of a tangent vector $X$ at $p$} is %can be written as
	\begin{eqnarray}
	\label{eqn:alpha_rep_tgt_vec}
	X_{\tilde{p}}^{(\alpha)}(x)
	& := & \left[\frac{{\tilde{p}}^{(\alpha)}(x)}{p(x)}X_{\tilde{p}}^{(e)}(x) - \frac{{\tilde{p}}^{(\alpha)}(x)}{p(x)}E_{{\tilde{p}}^{(\alpha)}}[X_{\tilde{p}}^{(e)}]\right]\nonumber\\
	& = & \left[\frac{p^{(\alpha)}(x)}{p(x)}\left(X_{\tilde{p}}^{(e)}(x) - E_{p^{(\alpha)}}[X_{\tilde{p}}^{(e)}]\right)\right],
	\end{eqnarray}
	where the last equality follows because ${\tilde{p}}^{(\alpha)} = p^{(\alpha)}$.
	The collection of all such $\alpha$-representations is
	\begin{eqnarray}
	T_{\tilde{p}}^{(\alpha)}(\mathcal{P}) := \{X_{\tilde{p}}^{(\alpha)} : X\in T_{\tilde{p}}(\mathcal{\tilde{P}})\}.
	\end{eqnarray}
	Clearly $E_{\tilde{p}}[X_{\tilde{p}}^{(\alpha)}] = 0$. Also, since any $A\in \mathbb{R}^{\mathbb{X}}$ with $E_{\tilde{p}}[A]=0$ is %can be written as
	\begin{eqnarray*}
		A = \left[\frac{p^{(\alpha)}}{\tilde{p}}\left(B-E_{p^{(\alpha)}}[B]\right)\right]
	\end{eqnarray*}
	with $B = \tilde{B}-E_{\tilde{p}}[\tilde{B}],$ where
	\[
	\tilde{B}(x) := \left[\frac{\tilde{p}(x)}{p^{(\alpha)}(x)} A(x)\right].
	\]
	In view of (\ref{exponential_tangent_space}), we have
	\begin{eqnarray}
	\label{e_space_equalto_alpha_space}
	T_{\tilde{p}}^{(e)}(\mathcal{\tilde{P}}) = T_{\tilde{p}}^{(\alpha)}(\mathcal{\tilde{P}}).
	\end{eqnarray}
	Now the inner product between any two tangent vectors $X,Y\in T_{\tilde{p}}(\mathcal{\tilde{P}})$ defined by the $\alpha$-information metric in (\ref{eqn:alpha-metric-bayesian}) is %can be written as
	\begin{eqnarray}
	\label{eqn:alpha_metric_general}
	\langle X,Y\rangle^{(\alpha)}_{\tilde{p}} := E_{\tilde{p}}[X^{(e)}Y^{(\alpha)}].
	\end{eqnarray}
	Consider now an $n$-dimensional statistical manifold $S$, a submanifold of $\mathcal{\tilde{P}}$, together with the metric $G^{(\alpha)}$ as in (\ref{eqn:alpha_metric_general}). Let $T_{\tilde{p}}^*(S)$ be the dual space (cotangent space) of the tangent space $T_{\tilde{p}}(S)$ and let us consider for each $Y\in T_{\tilde{p}}(S)$, the element $\omega_Y\in T_{\tilde{p}}^*(S)$ which maps $X$ to $\langle X,Y\rangle^{(\alpha)}$.  The correspondence $Y\mapsto \omega_Y$ is a linear map between $T_{\tilde{p}}(S)$ and $T_{\tilde{p}}^*(S)$. An inner product and a norm on $T_{\tilde{p}}^*(S)$ are naturally inherited from $T_{\tilde{p}}(S)$ by
	\[
	\langle \omega_X,\omega_Y\rangle_{\tilde{p}} := \langle X,Y\rangle^{(\alpha)}_{\tilde{p}}
	\]
	and
	\[
	\|\omega_X\|_{\tilde{p}} := \|X\|_{\tilde{p}}^{(\alpha)} = \sqrt{\langle X,X\rangle^{(\alpha)}_{\tilde{p}}}.
	\]
	Now, for a (smooth) real function $f$ on  $S$, the \emph{differential} of $f$ at $p$, $(\text{d}f)_{\tilde{p}}$, is a member of $T_{\tilde{p}}^*(S)$ which maps $X$ to $X(f)$. The \emph{gradient of $f$ at p} is the tangent vector corresponding to $(\text{d}f)_{\tilde{p}}$, hence, satisfies
	\begin{eqnarray}
	\label{eqn:differential_of_function}
	(\text{d}f)_{\tilde{p}}(X) = X(f) = \langle (\text{grad} f)_{\tilde{p}},X\rangle_{\tilde{p}}^{(\alpha)},
	\end{eqnarray}
	and
	\begin{eqnarray}
	\label{eqn:norm_of_differential}
	\|(\text{d}f)_{\tilde{p}}\|_{\tilde{p}}^2 = \langle (\text{grad}f)_{\tilde{p}},(\text{grad}f)_{\tilde{p}}\rangle_{\tilde{p}}^{(\alpha)}.
	\end{eqnarray}
	Since $\text{grad}f$ is a tangent vector, %we can write
	\begin{equation}
	\label{eqn:grad-f}
	\text{grad}f = \sum\limits_{i=1}^n h_i \partial_i
	\end{equation}
	for some scalars $h_i$. Applying (\ref{eqn:differential_of_function}) with $X = \partial_j$, for each $j=1,\dots,n$, and using (\ref{eqn:grad-f}), we obtain
	\begin{eqnarray*}
		(\partial_j)(f)
		& = & \left\langle \sum\limits_{i=1}^n h_i \partial_i, \partial_j\right\rangle^{(\alpha)}\\
		& = & \sum\limits_{i=1}^n h_i \langle \partial_i, \partial_j\rangle^{(\alpha)}\\
		& = & \sum\limits_{i=1}^n h_i g_{i,j}^{(\alpha)}, \quad j = 1, \dots, n.
	\end{eqnarray*}
	This yields
	\[
	[h_1,\dots,h_n]^T = \left[G^{(\alpha)}\right]^{-1}[\partial_1(f),\dots,\partial_n(f)]^T,
	\]
	and so
	\begin{equation}
	\label{eqn:grad-coeff-equation}
	\text{grad}f = \sum\limits_{i,j} (g^{i,j})^{(\alpha)}\partial_j(f) \partial_i.
	\end{equation}
	From (\ref{eqn:differential_of_function}), (\ref{eqn:norm_of_differential}), and (\ref{eqn:grad-coeff-equation}), we get
	\begin{eqnarray}
	\label{differential_and_metric}
	\|(\text{d}f)_{\tilde{p}}\|_{\tilde{p}}^2 = \sum\limits_{i,j} (g^{i,j})^{(\alpha)}\partial_j(f) \partial_i(f)
	\end{eqnarray}
	where $(g^{i,j})^{(\alpha)}$ is the $(i,j)$th entry of the inverse of $G^{(\alpha)}$.
	
	With these preliminaries, we %can 
	now state our main results. These are analogous to those in \cite[Sec.~2.5]{2000xxMIG_AmaNag}.
	\begin{theorem}
	\label{thm:variance_and_norm_of_differential}
		Let $A:\mathbb{X}\to\mathbb{R}$ be any mapping (that is, a vector in $\mathbb{R}^{\mathbb{X}}$. Let $E[A]:\mathcal{\tilde{P}}\to \mathbb{R}$ be the mapping $\tilde{p}\mapsto E_{\tilde{p}}[A]$. We then have
		\begin{eqnarray}
		\label{eqn:variance_and_norm_of_differential}
		\text{Var}_{p^{(\alpha)}}\left[\frac{\tilde{p}}{p^{(\alpha)}}(A-E_{\tilde{p}}[A])\right] =  \|(\text{d}E_{\tilde{p}}[A])_{\tilde{p}}\|_{\tilde{p}}^2.
		\end{eqnarray}
		$\hfill$% \QEDopen$
	\end{theorem}
	\begin{proof}
		For any tangent vector $X\in T_{\tilde{p}}(\mathcal{\tilde{P}})$,
		\begin{eqnarray}
		\label{eqn:tangent_acting_on_expectation}
		X(E_{\tilde{p}}[A])
		& = & \sum\limits_x X(x)A(x)\nonumber\\
		& = & E_{\tilde{p}}[X_{\tilde{p}}^{(e)} \cdot A]\\
		& = & E_{\tilde{p}}[X_{\tilde{p}}^{(e)}(A-E_{\tilde{p}}[A])].
		\end{eqnarray}
		Since $A-E_{\tilde{p}}[A]\in T_{\tilde{p}}^{(\alpha)}(\mathcal{\tilde{P}})$ (c.f.~(\ref{e_space_equalto_alpha_space})), there exists $Y\in T_{\tilde{p}}(\mathcal{\tilde{P}})$ such that $A-E_{\tilde{p}}[A] = Y_{\tilde{p}}^{(\alpha)}$, and $\text{grad}(E[A]) = Y$. Hence we see that
		\begin{eqnarray*}
		\lefteqn{\|(\text{d}E[A])_{\tilde{p}}\|_{\tilde{p}}^2}\\
		& = & E_{\tilde{p}}[Y_{\tilde{p}}^{(e)}Y_{\tilde{p}}^{(\alpha)}]\\
			& =  &   E_{\tilde{p}}[Y_{\tilde{p}}^{(e)}(A-E_{\tilde{p}}[A])]\\
			& \stackrel{(a)}{=} &\displaystyle E_{\tilde{p}}\left[\left\{\frac{\tilde{p}(X)}{ p^{(\alpha)}(X)} (A-E_{\tilde{p}}[A]) + E_{p^{(\alpha)}}[Y_{\tilde{p}}^{(e)}]\right\}(A-E_{\tilde{p}}[A])\right]\\
			& \stackrel{(b)}{=} & E_{\tilde{p}}\left[\frac{p(X)}{p^{(\alpha)}(X)}(A-E_{\tilde{p}}[A])(A-E_{\tilde{p}}[A])\right]\\
			& = &  E_{p^{(\alpha)}}\left[\frac{\tilde{p}(X)}{p^{(\alpha)}(X)}(A-E_{\tilde{p}}[A])\frac{\tilde{p}(X)}{p^{(\alpha)}(X)}(A-E_{\tilde{p}}[A])\right]\\
			& = &  \text{Var}_{p^{(\alpha)}}\left[\frac{\tilde{p}(X)}{p^{(\alpha)}(X)}(A-E_{\tilde{p}}[A])\right],
		\end{eqnarray*}
		where the equality (a) is obtained by applying (\ref{eqn:alpha_rep_tgt_vec}) to $Y$ and (b) follows because $E_{\tilde{p}}[A-E_{\tilde{p}}[A]] = 0$.
	
	\end{proof}
	
	\begin{corollary}
		\label{cor:variance_and_norm_of_differential_inequality}
		If $\tilde{S}$ is a submanifold of $\mathcal{\tilde{P}}$, then
		\begin{eqnarray}
		\label{variance_and_norm_of_differential}
		\text{Var}_{{p}^{(\alpha)}}\left[\frac{\tilde{p}(X)}{p^{(\alpha)}(X)}(A-E_{\tilde{p}}[A])\right] \ge \|(\text{d}E[A]|_{S})_{\tilde{p}}\|_{\tilde{p}}^2
		\end{eqnarray}
		with equality if and only if $$A-E_{\tilde{p}}[A]\in \{X_{\tilde{p}}^{(\alpha)} : X\in T_{\tilde{p}}(S)\} =: T_{\tilde{p}}^{(\alpha)}(S).$$ $\hfill$% \IEEEQEDopen$
	\end{corollary}
	\begin{proof}
		Since $(\text{grad }E[A]|_{S})_{\tilde{p}}$ is the orthogonal projection of $(\text{grad }E[A])_{\tilde{p}}$ onto $T_{\tilde{p}}(S)$, the proof follows from Theorem~\ref{thm:variance_and_norm_of_differential}.
	\end{proof}
	
	We use the aforementioned ideas to establish an $\alpha$-version of the Cram\'er-Rao inequality for the $\alpha$-escort of the underlying distribution. This gives a lower bound for the variance of an estimator of $S^{(\alpha)}$ starting from an unbiased estimator of $S$.

\section{Derivation of Error Bounds}
\label{sec:error_bounds}
We state our main result in the following theorem.

\begin{theorem} [Bayesian $\alpha$-Cram\'{e}r-Rao inequality]
	\label{thm:Bayesian_alpha_CRLB}
	Let $S = \{p_{\theta} : \theta = (\theta_1,\dots,\theta_m)\in\Theta\}$ be the given statistical model and let $\tilde{S}$ be as before. Let $\hat{\theta} = (\hat{\theta}_1,\dots,\hat{\theta}_m)$ be an unbiased estimator of $\theta = (\theta_1,\dots,\theta_m)$ for the statistical model $S$. Then
	\begin{equation}
	\label{eqn:Bayesian_alpha_cramerrao}
	   \int \text{Var}_{\theta^{(\alpha)}}\left[\frac{\tilde{p_\theta}(X)}{p_\theta^{(\alpha)}(X)}(\hat{\theta}(X) - \theta)\right] d\theta
	    \ge \left\{E_\lambda\big[G_\lambda^{(\alpha)}\big]\right\}^{-1},
	\end{equation}
	where $\theta^{(\alpha)}$ denotes expectation with respect to $p_{\theta}^{(\alpha)}$. (In (\ref{eqn:Bayesian_alpha_cramerrao}), we use the usual convention that, for two matrices $A$ and $B$, $A\ge B$ means that $A-B$ is positive semi-definite.)
	\end{theorem}

\begin{IEEEproof}
Given an unbiased estimator $\hat{\theta}$ of $\theta$ for $\tilde{S}$, let $A = \sum\limits_{i=1}^m c_i \hat{\theta_i}$, for $c = (c_1,\dots,c_m)\in \mathbb{R}^m$.

Then, from (\ref{variance_and_norm_of_differential}) and (\ref{differential_and_metric}), we have
	\begin{eqnarray}
	\label{eq:analogous_cramer_rao_inequality1}
	c\text{Var}_{\theta^{(\alpha)}}\left[\frac{\tilde{p_\theta}(X)}{p_\theta^{(\alpha)}(X)}(\hat{\theta}(X) - \theta)\right]c^t \ge c\{\lambda(\theta)G^{(\alpha)}_\lambda\}^{-1}c^t.
	\end{eqnarray}
	Integrating the above over $\theta$, we get
	\begin{eqnarray}
	\label{eq:analogous_cramer_rao_inequality2}
	c\int \text{Var}_{\theta^{(\alpha)}}\left[\frac{\tilde{p_\theta}(X)}{p_\theta^{(\alpha)}(X)}(\hat{\theta}(X) - \theta)\right] d\theta~c^t\nonumber\\
		\ge c~ \int [\lambda(\theta)G^{(\alpha)}_\lambda]^{-1} d\theta ~ c^t.
	\end{eqnarray}
	But
	\begin{equation}
	    \int [\lambda(\theta)G^{(\alpha)}_\lambda]^{-1} d\theta \ge \big\{\mathbb{E}_\lambda [G_\lambda^{(\alpha)}(\theta)]\big\}^{-1}
	\end{equation}
	by \cite{GrovesRothenberg1969Biometrika}. This proves the result.
\end{IEEEproof}	
\begin{remark}
\begin{enumerate}%[label=(\arabic*)]
\item[]
\item The above result reduces to the usual Bayesian Cramer-Rao inequality when $\alpha = 1$ as in \cite{kumar2018information}.
\item When $\lambda$ is the uniform distribution, we obtain the $\alpha$-Cramer-Rao inequality as in \cite{kumar2020cram}.
\item When $\alpha = 1$ and $\lambda$ is the uniform distribution, this yields the usual deterministic Cramer-Rao inequality.
\end{enumerate}
\end{remark}

\section{Conclusion}
\label{sec:conc}
We have shown that our Theorem~\ref{thm:Bayesian_alpha_CRLB} provides a general information geometric characterization of the statistical manifolds linking them to the Bayesian $\alpha$-CRLB for vector parameters; the extension to estimators of measurable functions of the parameter $\theta$ is trivial. We exploited the general definition of relative $\alpha$-entropy in the Bayesian case. This is an improvement over Amari’s work \cite{amari2000methods} on information geometry which only dealt with the notion of deterministic CRLB of scalar parameters. Further, this is a generalization of our earlier information-geometric frameworks of Bayesian \cite{kumar2018information} and deterministic $\alpha$-CRLB \cite{kumar2020cram}. These improvements enable usage of information geometric approaches for biased estimators and noisy situations as in radar and communications problems \cite{mishra2017performance}. 

%% References:
%\clearpage
%\balance
\bibliographystyle{IEEEtran}
\bibliography{main}

\end{document}